\documentstyle[11pt,newpasp,twoside,epsf]{article}
\def\edcomment#1{\iffalse\marginpar{\raggedright\sl#1\/}\else\relax\fi}
\reversemarginpar

\begin{document}
\title{$^7$Li in Metal-Poor Stars: The Spread of the Li Plateau}
 \author{Sean G. Ryan}
\affil{Dept of Physics and Astronomy, The Open University, Walton Hall, Milton Keynes MK7 6AA, UK}

\begin{abstract}
A highly homogeneous study of 23 halo field dwarf stars has achieved a Li
abundance accuracy of 0.033~dex per star. The
work shows that the intrinsic spread of the Li abundances of these stars 
{\it at a given metallicity} is
$<~0.02$~dex, and consistent with zero. That is, the Spite Li plateau for
halo field dwarfs is incredibly thin. The thinness rules out
depletion by more than 0.1~dex by a rotational-induced extra-mixing mechanism.
Despite the thinness of the plateau, an
increase of Li with [Fe/H] is seen, interpreted as 
evidence of Galactic chemical evolution (GCE) of Li, primarily due to 
Galactic cosmic ray (GCR) spallation reactions in the era of halo formation.
The rate of Li evolution is concordant with: (1) observations
of spallative $^6$Li in halo dwarfs; (2) GCE models; and (3) data
on Li in higher metallicity halo stars. New data have also revealed four 
new ultra-Li-deficient halo dwarfs, doubling the number known. Based on their
propensity to cluster at the halo main sequence turnoff {\it and also} to
exist redward of the turnoff, we hypothesise that they are the products of 
binary mergers that ultimately will become blue stragglers.
We explain their low Li abundances by normal 
pre-main-sequence (and possibly main-sequence) destruction in the
low mass stars prior to their merging. If this explanation is correct,
then such stars need no longer be considered an embarrassment to
the existence of negligible Li destruction in the majority of
field halo dwarfs.

\end{abstract}

\section{Introduction}

The first indication that the old stars of the Galaxy exhibited an almost
uniform Li abundance emerged at IAU Coll. 68, when Spite \& Spite (1981, 1982)
presented their first observations of warm halo dwarfs. Almost two decades 
later, IAU Symp. 198 met to consider progress
in studies of this and other light elements.

Studies by many workers in the decade following the Spite \& Spite discoveries 
resulted in mounting evidence that the warm halo dwarfs exhibited a unique Li 
abundance (e.g. Rebolo, Molaro, \& Beckman 1988). The interpretation of this 
abundance as the primordial one reflecting big bang nucleosynthesis, at
worst ``hardly altered'' (Spite \& Spite 1982), hinged on the  importance of
possible depletion of Li from a higher initial abundance. 
While ample
evidence existed of Li destruction in some stars, the lack of a
significant spread in halo dwarf Li abundances provided empirical evidence that
destruction may have been minimal in these objects. (See Boesgaard \& Steigman
1985 for a review of that period.) Classical stellar evolution models
(e.g. Deliyannis, Demarque, \& Kawaler 1990) fitted this interpretation,
showing that Li destruction in metal-deficient dwarfs with shallow surface 
convective zones would be minimal ($^<_\sim$ 0.05~dex). However, this same
class of stellar
models failed to explain numerous Population I star observations, and an
alternative class of models invoking extra mixing implied that considerable
Li depletion (as high as 1 dex; Pinsonneault, Deliyannis, \& Demarque 1992) 
could have occurred in the halo stars.

In the next decade, several dissenting voices were heard.
Deliyannis, Pinsonneault, \& Duncan (1993) argued that there was a 
non-negligible spread in the Li abundances of the halo dwarfs that would not
be consistent with a perfectly primordial composition. Depending on the
sample selected, they found a Li spread of $\sigma~\ge~0.04$~dex.
Thorburn (1994) found an even greater intrinsic spread 
$\sigma~\simeq~0.10$~dex, and moreover claimed, as did Norris, Ryan, \& 
Stringfellow (1994), that the abundances depended on both $T_{\rm eff}$ and
[Fe/H]. Such dependences were contrary to the notion of a unique Li 
abundance in the halo stars, and thus undermined the association of the
observed Li abundance(s) with the primordial one. The efforts of 
Ryan et al. (1996) to bring all previous observations onto a uniform temperature
and abundance scale did not eliminate the cited dependences.

One of the largest uncertainties in abundance analyses is errors
in the effective temperature scales. Spite \& Spite (1993) and 
Bonifacio \& Molaro (1997) discussed the possible role of such errors in
distorting an otherwise uniform Li abundance, the latter work finding the
previously reported trends to become insignificant when a more recent 
temperature calibration based on the infra-red flux method (IRFM) was applied.
Although the IRFM scale might be expected to provide better systematics,
the large individual uncertainties attached to each temperature determination 
by this method limit the scale's ability to distinguish between effects at the
level of those claimed for Li. The existence of some large errors even in the
metallicity estimates for program stars also hampers the efforts. In
particular, the literature data utilised by Bonifacio \& Molaro (1997) includes
several poorly determined values which, in hindsight, frustrated their analysis 
by smearing out the data (Ryan, Norris, \& Beers 1999, \S7.3.3).

\section{The Intrinsic Spread of $^7$Li}

In an effort to avoid the impact of undesirable errors, Ryan, Norris, \& Beers
(1999) set out to obtain a highly homogeneous data set on a sample occupying 
only a
narrow range of $T_{\rm eff}$, [Fe/H], and evolutionary type. Restricting their
sample to 6000~K~$^<_\sim~T_{\rm eff}~^<_\sim~$~6400~K and
$-3.5~^<_\sim$~[Fe/H]~$^<_\sim~-2.5$, applying double-blind data analysis
techniques, obtaining multiple high-resolution, high-S/N observations of 
the targets, and using multiple temperatures indicators to minimise random
errors, 
they achieved a formal abundance error as low as $\sigma_{\rm err}$~=~0.033~dex
per star. These results are at considerably higher precision than most previous
Li measurements (typically having $\sigma_{\rm err}$~$\simeq$~0.06 -- 0.08 dex).

The sample was known to contain one previously known ultra-Li-deficient star,
G186-26, which was excluded from the analysis. Remaining objects exhibited
a total observed spread $\sigma_{\rm obs}$~=~0.053~dex, considerably less than 
that found by 
Thorburn (1994). However, this 0.053~dex was found to be dominated
by an underlying metallicity dependence, and 
the spread of the Li abundances about this trend is a mere 
$\sigma_{\rm obs}$~=~0.031~dex (see Figure 1), and Gaussian in form. 
This corresponds to the spread in Li abundance
at a given metallicity. Comparing this with the formal measurement errors of
$\sigma_{\rm err}$~=~0.033~dex leads to the conclusion that the intrinsic
spread in the stars must be negligible. We state a generous upper-limit on the
intrinsic spread as being $\sigma_{\rm int}~<~0.02$~dex.

\begin{figure}
\plotfiddle{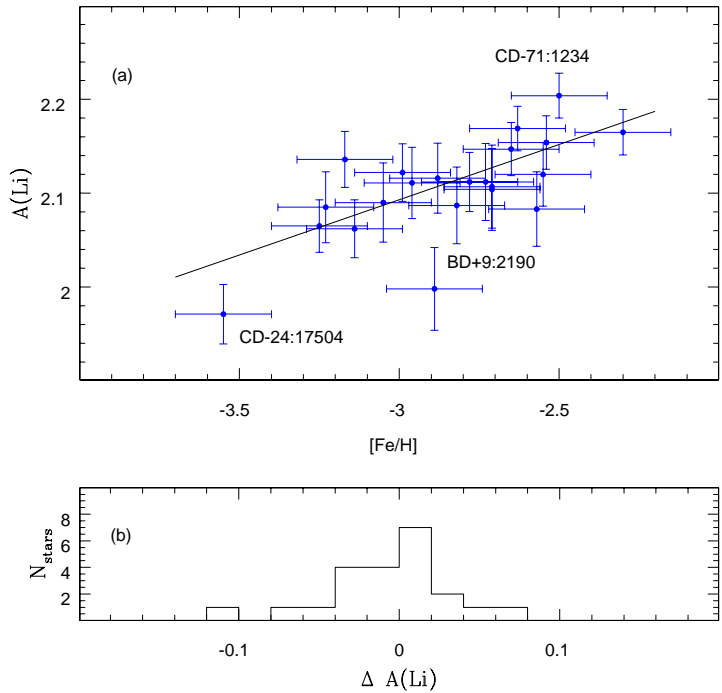}{080mm}{0}{50}{50}{-150}{-120}
\caption{Metallicity-dependence of turnoff Li abundances, and residual
observed spread with $\sigma_{\rm obs}$~=~0.031~dex.}
\end{figure}

An important consequence of the very narrow spread of Li abundances is its
ability to constrain the impact of possible extra-mixing in so far as
extra-mixing models predict a spread in the final Li abundances of a population
of stars.
The rotationally-induced mixing models of
Pinsonneault et al. (1993) suggested that Li depletion by as much as
an order of magnitude could have occurred in halo turnoff
dwarfs. The more recent work of Pinsonneault et al. (1999), in concert with the
observational data of Thorburn (1994), revised downward the
depletion level to $\simeq 0.2$~--~0.4~dex. As Figure~2 shows, the data from
Ryan et al. (1999) with their narrower spread (at a given metallicity)
rule out rotationally-induced mixing models that exhibit even 0.1 dex median
depletion. Considering the size of the observed sample and the absence of stars
in the tail of the theoretical 
distribution, Poisson statistics provide only 10\% chance that the observed and
theoretical curves are compatible,
or 90\%\ probability that the median
depletion is less than 0.1~dex. Other statistical tests 
discriminate the models even more significantly.  

\begin{figure}
\plotfiddle{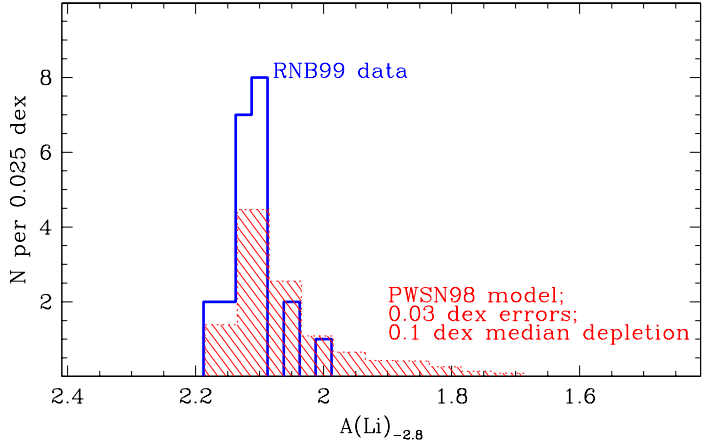}{060mm}{0}{50}{50}{-150}{-190}
\caption{Spread in observations (at a given metallicity after compensation for 
the [Fe/H] dependence of Li), compared with predictions for a 
rotationally-induced mixing model exhibiting a median depletion of 0.1~dex.}
\end{figure}

We seek now to explain
previous results that yielded contrary conclusions. 
Thorburn's (1994) data reduction process explicitly excluded sky background and 
scattered light subtractions, but this shortcut was not reflected in the
formal error estimates. Incorporation of the errors introduced by this procedure
are enough, on average, to inflate the error estimates to the size required by
the observed scatter. That is, the scatter observed by Thorburn is almost
certainly consistent with that resulting from data acquisition and analysis.
Bonifacio \& Molaro (1997) found no significant metallicity dependence in their
analysis, but as discussed above, certain metallicities they adopted from the 
literature were found subsequently to be unreliable. This, and the large
random errors inherent in the IRFM temperature scale, resulted in the weak
Li evolution being washed out. (See Ryan et al. 1999, \S7.3.3, for a detailed
analysis.) Finally, we note that the small spread of
abundances found by Deliyannis et al. (1993) is consistent with the observed
spread in our sample if the underlying metallicity trend is overlooked.
In fact, the Deliyannis et al. study pre-dated any claims of a metallicity
dependence, and their result was probably driven by the large metallicity
range in their sample.

\section{The Underlying Li vs [Fe/H] Trend}

Although Li GCE during the halo-forming era has often been ignored, we should 
not be surprised that it exists. If recent
detections of $^6$Li in halo stars (Smith, Lambert, \& Nissen 1993,1998;
Hobbs \& Thorburn 1994,1997; Cayrel et al. 1999, Deliyannis \& Ryan 2000) 
are correct, then we
would be surprised {\it not} to see $^7$Li GCE. With the measurement
precision attainable using modern CCDs and large aperture telescopes, 
even small levels of $^7$Li enrichment can be measured, and it is
consistent with the measured $^6$Li abundances.

To see whether the observed trend was compatible with GCE,
Ryan et al. (2000a) examined the Fields \& Olive (1999a,b) model.
For halo stars, the $\nu$-process and GCR spallation are the most
likely sources of $^7$Li. The Fields \& Olive model normalises the GCR
contribution to meteoritic Be and $^6$Li abundances, and 
normalises the $\nu$-process to the otherwise unaccounted for $^{11}$B.
The model does not include stellar $^7$Li sources acting in the later stages of
Galactic evolution, and hence does not model the Population I abundance.
The models of Romano et al. (1999) incorporate many Population I sources
(primarily the $\nu$-process, AGB stars, and novae).
Figure~3 shows two variants of Romano et al's models and a hybrid
using the GCR predictions of Fields \& Olive from 
Ryan et al. (2000a). The model reproduces not only the halo star Li evolution 
discussed above, but also fits new
data around [Fe/H]~$\sim$~$-1.5$ (Ryan et al. 2000b; see below)
and the lowest metallicity datum at [Fe/H]~=~$-3.7$ (Norris, Beers, \& Ryan 2000)
which were added {\it after} the Fields \& Olive model was produced.

\begin{figure}
\plotfiddle{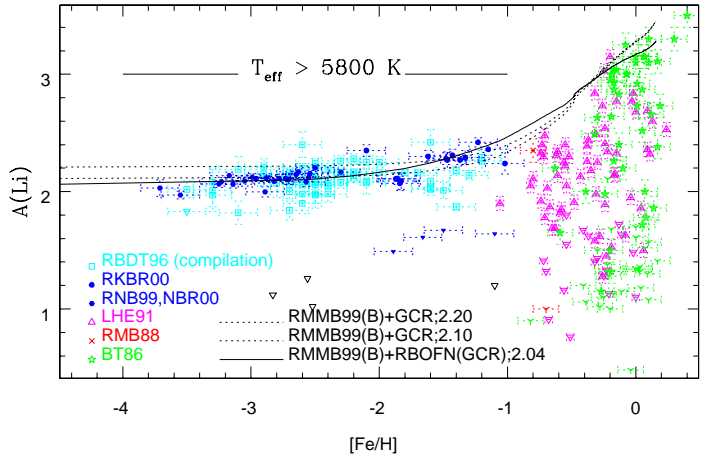}{060mm}{0}{50}{50}{-150}{-190}
\caption{Evolution of Li with metallicity. Observations are for halo stars
having $T_{\rm eff}~>~5800$~K, to avoid lower-mass stars with Li
depletion and to reduce the heterogeneity of the sample, and Population I stars
from sources indicated. Models are (dashed curves) from 
Romano et al. (1999) for two different
primordial values ($A$(Li)$_p$~=~2.10 and 2.20), 
and (solid curve) a hybrid model using the GCR contribution of Fields \&
Olive (1999a,b; Ryan et al. 2000a) with Population I evolution from 
Romano et al. }
\end{figure}

Ryan et al. (1996) indicated that a selection bias existed in the Li data
available in the literature, in that most studies
centred on more-metal-poor objects. Few more-metal-rich
halo stars had been examined, and those which had were on the whole cooler than
the metal-poor ones. Given the difficulties with
temperature scales for stars, the comparing of warmer metal-poor stars with
cooler metal-rich stars was clearly undesirable. In an effort to address this
bias, Ryan et al. (2000b) obtained data on 18 more-metal-rich halo stars, with 
$-2.0~^<_\sim$~[Fe/H]~$^<_\sim~-1.0$, but in the warm temperature
range $T_{\rm eff}~^>_\sim~6000$~K as for the most metal-poor samples.
The sample was found to contain four ultra-Li-deficient halo stars, which will
be discussed separately in the next section. The remaining stars, shown as
solid circles in Figure~3, were found to sit exactly where the Fields \&
Olive model predicts. We emphasise that the model was completed 
prior to the reduction and analysis of the metal-rich halo sample, so the
agreement between the two is a genuine accomplishment, not something achieved
artificially in the model. This is viewed as 
additional evidence that the trend of Li with [Fe/H] evidenced in Figure~1
is a result of natural GCE of the element during formation of the 
Galactic halo.

\section{The Primordial Li Abundance and Uncertainties}

We combine these measurements
in Table~1 to present a new accounting for the primordial Li abundance.
Beginning with the observed abundance at the mean metallicity of our sample,
we apply corrections for the inferred GCE contribution (with uncertainties)
and for stellar depletion. For the latter we take the value implied by
classical models, but in the uncertainties allow for additional depletion
up to the 0.1~dex limit of the rotationally-induced mixing models. 
Temperature scale uncertainties remain one of the largest sources of error,
and in this analysis we apply an offset of 0.08~dex to the Li abundance,
corresponding to a change from the temperature scale adopted in our original 
analysis (based on the Bell \& Oke (1986) and Magain (1987) b-y scales) to
the systematically hotter IRFM scale of 
Alonso, Arribas, \& Mart\'inez--Roger (1996).
However, we associate an uncertainty
of $\pm$0.08~dex with this process, in recognition of the remaining 
difficulties in the temperature scales for halo dwarfs.
These and the other affects tabulated lead us to infer a primordial abundance
$A$(Li)$_p$~=~2.09~$^{+0.19}_{-0.13}$~dex, where the uncertainties resemble
2$\sigma$ limits (Ryan et al. 2000a).

\begin{table}[htb]
\begin{center}
\caption{Transforming the observed halo Li abundance into the primordial 
abundance, accounting for random and systematic errors.}
\begin{tabular}{lll}
\hline
\multicolumn{3}{l}{\bf Systematic Effects Influencing Inferred} \\
\multicolumn{3}{c}{\bf Primordial Lithium Abundance}\\
\hline
Observed:\\
\multicolumn{1}{r}{$\langle A$(Li)$_{-2.8}\rangle$ =}
						&2.12   &$\pm$0.02\\
\hline
\multicolumn{3}{l}{Corrections to apply:}\\
GCE/GCR                                         &$-0.11$&$^{+0.07}_{-0.09}$\\
Stellar depletion                               &$+0.02$&$^{+0.08}_{-0.02}$\\
$T_{\rm eff}$ scale zeropoint                   &$+0.08$&$\pm$0.08\\
1-D atmosphere models                           &$+0.00$&$^{+0.10}_{-0.00}$\\
Model temperature gradient                      &$+0.00$&$^{+0.08}_{-0.00}$\\
NLTE                                            &$-0.02$&$\pm$0.01\\
$gf$-values                                     &$+0.00$&$\pm$0.04\\
Anomalous/pathological objects                  &$+0.00$&$\pm$0.01\\
\ \\
Total                                           &$-0.03$&$^{+0.19}_{-0.13}$\\
\hline
Inferred:\\
\multicolumn{1}{r}{$A$(Li)$_p$ =}               &2.09   &$^{+0.19}_{-0.13}$\\
\hline
\end{tabular}
\end{center}
\end{table}

\section{The Ultra-Li-Deficient Halo Dwarfs: Blue Stragglers After All?}

Boesgaard \& Tripicco (1986a) showed that Hyades stars with
6400~K~$<~T_{\rm eff}~<$ 6900~K exhibit extremely low surface Li abundances.
These and similar stars in other Population I
clusters became known as ``Li-dip stars''. They showed that a second mechanism, 
besides convection on the pre-main sequence (and possibly main sequence) for
lower mass stars, could greatly deplete surface Li abundances. 
Lambert, Heath, \& Edvardsson (1991) showed that {\it most} of the strongly 
Li-depleted
Population I field stars, shown for example in Figure~3, could be explained 
as either having evolved from the Li dip or being low mass
convectively-depleted stars. However, they also noted that a small number of
stars did not share these histories, and proposed that perhaps 10\%\ of 
stars had experienced additional severe Li depletion through unknown
causes. This work preceded the discovery of halo dwarfs whose temperatures and
metallicities coincided with the Spite plateau but which were 
ultra-Li-deficient by more than an order of magnitude or so (Hobbs, Welty
\& Thorburn 1991; Thorburn 1992; Spite et al. 1993). The halo examples
have been estimated at perhaps 3--5\%\ of the Population, and may result from 
the same process as the Population I class proposed by Lambert et al.
The nature of the process resulting in their Li-depletion has remained 
unclear, and our inability to explain them has been given as a reason to
mistrust the entire Population II interpretation (e.g. Thorburn 1994).
Whether or not such a view is held (cf. Ryan et al. 1999), they identify an
embarrassing deficit in our knowledge of stellar processing of this
important element. Efforts to identify common chemical signatures other than
Li deficiency proved impossible; instead considerable diversity and
heterogeneity was found
amongst the complete sample (four) known at the end of 1998 
(Norris et al. 1997; Ryan, Norris, \& Beers 1998).

The study of 18 more-metal-rich halo stars by Ryan et al. (2000b) discussed
above resulted in the discovery of four new ultra-Li-deficient halo stars;
see Figure~4
(Ryan et al. 2000c). The discovery rate in that study, 22\%, contrasts
greatly with the previous Population estimate of 3--5\%, and indicates that
they are preferentially clustered in the stellar parameter range singled out
in that investigation, namely warm, more-metal-rich halo stars. The stars are
therefore seen to be grouped preferentially towards the main sequence turnoff 
of the halo, but not exclusively so. The clustering near the turnoff is
reminiscent of blue stragglers, but the hypothesis that they were
redward-evolving blue stragglers had already been ruled out for the 
previously known examples (Thorburn 1994; Norris et al. 1997).
However, the discovery of four more such stars preferentially close to the 
main sequence turnoff resulted in the re-examination of the blue-straggler
hypothesis, with the distinction that main-sequence blue-stragglers-to-be
are now considered. 
Halo stars initially cooler than about
5700~K, corresponding to a mass of $\sim~0.7~M_\odot$, deplete their
surface Li during their pre-main-sequence evolution. 
When two low-mass stars merge to become a single object higher up the 
main sequence near the halo turnoff with a total mass around
0.7-0.85~$M_\odot$, they will in most cases form from stars which have already
destroyed their Li. They will appear, then, as normal halo main-sequence stars
{\it except} with respect to two parameters: (1) their Li abundances will be
extremely low and hence appear abnormal, and (2) their main sequence
lifetimes will be extended due to the delayed onset of nuclear burning at a 
rate expected of stars of their (combined) mass. The stars we now observe as
ultra-Li-deficient halo stars, preferentially but not uniquely clustered towards
the turnoff, may indeed be the progeny of such mergers and the progenitors of
future blue-stragglers. Indeed, extreme Li deficiency may be the
only common signal of future Pop II blue-stragglers-to-be.

\begin{figure}
\plotfiddle{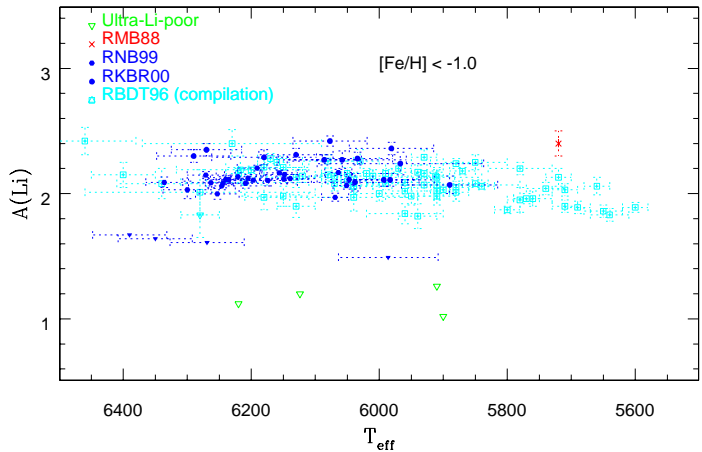}{060mm}{0}{50}{50}{-150}{-190}
\caption{Recent observations doubling the number of known
ultra-Li-deficient halo stars. Their location 
close to the main sequence turnoff leads us to the hypothesis
that they are the progeny of low-mass binary-star 
mergers, destined to become blue-stragglers.}
\end{figure}

If this hypothesis is correct, then we may finally remove such stars with
confidence from discussions of the spread about the Spite plateau, and
consider them as a truly distinct class of stars whose evolutionary history
explains their abnormal Li abundances. Whether this mechanism can also
explain the heterogeneity found for the abundances of their other elements
remains to be seen.

\section{Differences Between Halo Field and Globular Cluster Stars?}

Although the halo field stars discussed above have
minimal intrinsic spread about the Li Spite plateau (at a given abundance), 
data for globular cluster samples  show a different picture. Figure~5 compares
the very metal-poor field turnoff dwarf data for stars spanning a dex in [Fe/H]
with observations of subgiants in M92 (Boesgaard et al. 1998). The two groups
exhibit quite different Li characteristics!
The considerable spread in the globular cluster sample prompted Boesgaard et al.
to favour a mechanism in which a higher pre-stellar abundance has been
depleted by varying degrees in the stars, possibly by the rotationally-induced
mixing mechanism discussed earlier. Why this mechanism should differ for the
globular cluster and field star samples is unclear. Differences in
angular momentum evolution in the two environments may be responsible,
but the details have yet to be proposed. Other examples
of differences in the mixing of stellar envelopes in field star and globular
cluster samples has been forthcoming in recent years (e.g. Hanson et al. 1998),
adding to previous evidence of field-vs-cluster differences in
CNO element ratios.

\begin{figure}
\plotfiddle{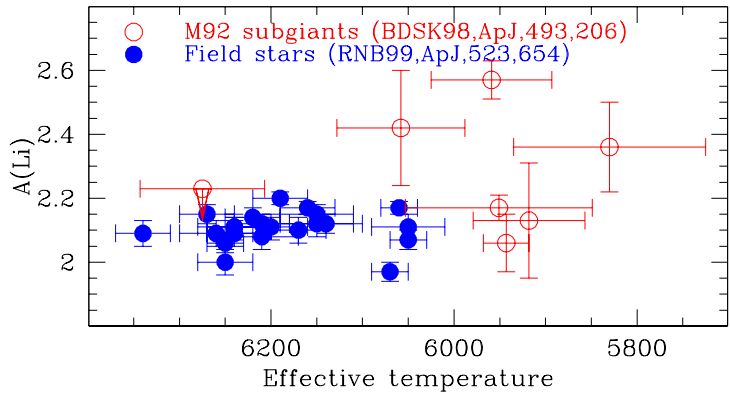}{050mm}{0}{50}{50}{-150}{-190}
\caption{Contrast between the tight distribution of Li for halo field dwarfs 
(having a range of [Fe/H]), compared with the broad spread in globular cluster 
subgiants. See text for discussion.}
\end{figure}

Until the cause of the difference is understood, one must choose whether to
use the field star or the globular cluster data to interpret GCE. I would
argue that the large Galactic volume sampled by field stars in contrast to
the small total volume of globular clusters, and the greatly increased 
possibility of star-to-star interactions in the high stellar densities of
the latter, would render field star samples more representative of the evolution
of the Galaxy as a whole. Of course, this in no way reduces the importance of 
understanding the globular cluster element abundance patterns for what they
may tell us about the evolution of the Galaxy and stellar processes in dense
environments.

\acknowledgments

This work represents the outcomes of collaborations involving
Prof. J. E. Norris (Australian National University),
Dr T. C. Beers (Michigan State University),
Dr K. A. Olive (University of Minnesota),
Dr B. D. Fields (University of Illinois at Urbana-Champaign),
Dr T. Kajino (National Astronomical Observatory of Japan),
Ms. D. Romano (SISSA, Trieste),
and
Ms K. Rosolankova (The Open University, \& St Hilda's College, Oxford), 
all of whose contributions are gratefully acknowledged.
The author is likewise grateful to the IAU for partial support to attend this
meeting.

\end{document}